\documentclass[12pt,longnamesfirst,preprint]{aastex}

\slugcomment{2010 Asia-Pacific Radio Science Conference\\
in Toyama, Japan on September 22-26, 2010\\
Commission J: Radio Astronomy\\
J2 Millimeter- and Sub-millimeter-wave Telescope and Array
}

\shorttitle{Saito et al.}
\shortauthors{
Next Generation Array
}

\begin{document}


\title{
Next Generation Millimeter/Submillimeter Array to Search for 2nd Earth
}


\author{
Masao Saito\altaffilmark{1}
 and Satoru Iguchi\altaffilmark{1}}  


\altaffiltext{1}{
National Astronomical Observatory of Japan, 2-21-1 Osawa,
Mitaka, Tokyo 181-8588, Japan
}




\section{Purpose} 
ALMA[1] is a revolutionary radio telescope at present and its full operation will start from 2012. It is expected that ALMA will resolve several cosmic questions and will show a new cosmic view to us. Our passion for astronomy naturally goes beyond ALMA because we believe that the 21st-century Astronomy should pursue the new scientific frontier. In this conference, we propose a project of the future radio telescope to search for Habitable planets and finally detect 2nd Earth as a Migratable planet. The detection of 2nd Earth is one of ultimate dreams for not only astronomers but also people. 

\section{Conceptual Design and Primary Scientific Requirements}
To directly detect 2nd Earth, we have to carefully design the sensitivity and angular resolution of the telescope by conducting trade-off analysis between the confusion limit and the minimum detectable temperature. Assuming an array that has 64 50-m antennas with 25 $\micron$ surface accuracy located within the area of 300 km, dual-polarization SSB receivers with noise temperature that is the best or improved performance of ALMA, and IF bandwidth of 128 or 256 GHz, the result of the sensitivity analysis is derived (see \ref{fig:fig1})[2]. We temporally name this telescope "Very Large Millimeter/Submillimeter Array (VLMSA)". Since this sensitivity is extreme high, we have a lot of chances to study the galaxy, star formation, cosmology and of course the new scientific frontier. 
	The primary science requirement for VLMSA is the flexibility to support the breadth of scientific investigation to be proposed by its creative scientist users over the decadefs long lifetime of the instruments. However, three science requirements stand out in all the science planning for VLMSA. These three level-1 primary science requirements are the following: 1) The ability to directly detect thermal radiation from 2nd Earth in other planetary systems like our solar system at a distance of 2 pc, in less than 24 hours of observation; 2) The ability to clearly image a black hole with an accretion disk in the active central region of Sagittarius A and M87; and 3) The ability to identify the absolute position of stars like Sun at a distance up to 1 kpc for the astrometry. 
From the viewpoints of sensitivity, the scientific requirements of 2) and 3) can be easily realized if meeting the requirement of 1). To resolve a black hole with an accretion disk in the radio core, the angular resolution of 20 $\mu$ arcsec may be necessary[3]. We will find a good solution to keep the extremely high sensitivity and to directly image a black hole with the angular resolution by locating a few antennas at up to 3000 km.


\end{document}